# Self-Healing by design:
# Immune-system inspired autonomous cyber resilience

**Dr. Shari Finner**
TNO Cyber Security Technologies
Eindhoven, The Netherlands
shari.finner@tno.nl

**Swarna Kumarswamy-Das**
TNO Cyber Security Technologies
The Hague, The Netherlands
swarna.kumarswamy@tno.nl

**Bart Gijsen**
TNO Cyber Security Technologies
The Hague, The Netherlands
bart.gijsen@tno.nl

## Abstract

**Digital innovations have strongly improved ship capabilities, including command, control & communication, navigation and combat management. In the wake of digitisation and geopolitical stress, resilience against cyber-attacks has become a key challenge. Autonomous Cyber Resilience (ACR) methods, inspired by features of the human immune system, provide a promising perspective. Proofs of concept of autonomous self-healing and self-protecting software have been realised in cloud environments using containerisation, monitoring and deployment technology. However, developing a suitable "cyber immune system" for more complex digital systems is still an intricate challenge.**

**In this paper, we share insights on how ACR methods can be combined with resilience mechanisms present in critical naval and aeronautical manoeuvring systems. Using TNO's "OpPic" application, we show how TNO's Self-Healing for Cyber Security software can provide autonomous mission- and threat-adaptive resilience. On this basis, six guidelines for the design of autonomous cyber resilient systems are derived. The ultimate goal is to enable operators to oversee and control their system's "cyber immune response", to be able to continuously adapt to evolving threats and mission priorities.**

## The impact of cyber-attacks

In times of digitalisation, Military, Air and Space Technologies are becoming more and more dependent on digital services. The growing amount of connected digital components and systems strongly increases the attack surface for digital threats such as cyber-attacks, system faults, misconfigurations as well as human errors. Meanwhile, we face a worldwide continuous rise of cyber-attacks, a significant part of which can be attributed to nation state actors aiming for espionage or sabotage opportunities. [1] It is becoming increasingly necessary both for civil and military organisations to become inherently cyber-resilient and stay ahead of (or at least keep on par with) the attackers.

Cyber resilience is the ability for a digital system to anticipate, withstand, recover from, and adapt to cyber-attacks, to ensure that system behaviour adheres to the business and operational objectives to the maximum possible extent. As cyber-attacks are often executed automatically at machine-speed and system compromises cannot always be prevented, cyber resilience requires the development of techniques for digital systems to not only protect themselves, but also recover automatically (or autonomously) in *real-time*, maintaining business- and mission continuity. [2] In the words of the UK Royal Navy: "*Systems must be resilient so that they can "take a punch" and fall-back to reversionary modes if needed whilst maintaining operational output*". [3]

The impact of cyber-attacks on mission critical systems can be severe: attacks on naval or air traffic navigation systems may lead to course deviations, crashes, or to the implementation of strict emergency speed- or capacity limits or shutdowns in order to avoid physical harm. Phishing emails can lead to malware affecting a ship's bridge systems or ballast. [4] Attacks on steering or propulsion systems may render vessels unusable, while attacks on mission-critical battlefield management systems, command-and-control systems physical defence systems or weapon systems may result in shutdown or even unintended (adversary) use. [5] According to Eurocontrol, "*the manipulation of satellite navigation signals has become mainstream*". [6] In a spoofing incident in 2021, NATO warships were (unrightfully) made to appear as if they were near Russian-occupied Crimea, entering Russia's main naval base, despite not actually being there. [7] As such, cyber incidents can have vast impacts, including the potential to provoke geopolitical escalation. However, only 31% of surveyed transportation, cargo and logistics businesses are planning to invest further in cyber security. [4]

To maintain mission continuity and societal stability, it is imperative to make any critical infrastructure and mission-critical technology resilient to cyber-attacks. This applies to communication systems and networks, but also to any connected IT and OT components such as radar and navigation systems. Cyber resilience is becoming a functional requirement instead of a nonfunctional that can be retrofitted, which was recently acknowledged also by the European Union through the publication of the Cyber Resilience Act. [8] This paper focusses on the step beyond cyber resilience: the design and development of *autonomous* cyber resilience (ACR), which is the ability of a (digital computing) system to anticipate, withstand, recover from, and adapt to cyber-attacks and failures to ensure mission or business objectives without being controlled directly by humans.

## Cyber Resilience State of the Art

In the current state of the art, a large amount of products and digital systems are not (yet) developed with any cyber resilience considerations in mind. Several cyber security guidebooks do exist for the fields of Air Transport management or cyber-physical vehicles. [9] However, these are applicable only to specific sectors, and solely safeguard a minimal baseline level of cyber security. To address the need for more cyber resilience in society, the Cyber Resilience Act (CRA) was adopted in 2024 – a European regulation aiming to broadly enforce cyber resilience compliance of products and systems in the European Union. [8] However, we expect the measures of the CRA to be insufficient for highly targeted organisations, and more so in times of advancing attack capabilities and geopolitical tensions.

Even before the CRA, a number of cyber security frameworks have been published, for example by MITRE and NIST to provide guidance for developing inherently cyber resilient systems. [10] These frameworks present several atomic resilience techniques such as monitoring, privilege restriction and segmentation, but again leave the implementation of a technique for specific operational cases up to the ingenuity of system developers and human operators. While very useful in their own right, these frameworks provide little concrete guidance on how to engineer integral machine-speed cyber resilience of complex systems.

Next to several cyber security frameworks, the concept of "autonomic computing" was introduced by IBM engineers in order to facilitate the task of monitoring, controlling and maintaining increasingly complex digital systems. [11] Their framework for self-management includes system requirements for self-configuration, self-healing, self-optimisation and self-protection of digital systems. As part of this concept, also the generic engineering techniques called MAPE was introduced. It includes control loops that integrate the actions of **M**onitoring, **A**nalysis, **P**lanning and **E**xecution, and is conceptually similar to the OODA loop (**O**bserve, **O**rient, **D**ecide, **A**ct), which is often used in military context. Until today, MAPE and OODA are very useful techniques for integrating monitoring and control concepts that can potentially be used for the design and automation of integral cyber resilience.

Interestingly, many resilience concepts, such as those of NIST and IBM's autonomic computing, have drawn inspiration from defensive properties of the immune system: The battle of the body against pathogens (viruses, bacteria, fungi and parasites) and "survival of the fittest" concept in biology shows strong parallels with the cyber security battle between attackers and defenders, leading to survival of the 'cyber-fittest' in dynamic and evolving digital environments. This parallel has given rise to several bio-inspired autonomous cyber resilience and cyber defence concepts in the past years.

**Autonomous Cyber Resilience (ACR)** is the ability of a digital system to protect, defend, recover and heal itself at machine-speed in response to dynamic changes in its environment, and without being controlled directly by humans. Similar to the human immune system, such a system employs protection and defence mechanisms continuously, proactively and independently by implementing redundancy and disposability, and by correcting any type of anomalous behaviour quickly, irrespective of it being classified as "malicious". This is in strong contrast to the reactive post-analysis and mitigation of already-felt impact, which is typical in traditional cyber security operations.

Internationally, a handful of organisations are starting to embrace concepts of self-protection, self-healing and autonomous defence against cyber-attacks. One commercial example is the "DEFENSA" solution by Akimbo technologies, which offers corrections (“healing”) of anomalies in military vehicle network communication subject to man-in-the-middle attacks. [5] The solution is based on dynamic distributed fault-tolerant computing on the datalink layer, implementing redundancy by triplicating network communication packages and subsequently enforcing transmission over disjoined communication paths. [12] Further examples are Darktrace’s “Enterprise Immune system” and “Antigena” solutions, which leverage AI-based decision making to automatically neutralise any abnormal behaviour in enterprise networks.

A recent cyber resilience solution is also provided by Dutch not-for-profit organisation TNO, with the open source Self-Healing for Cyber-Security (SH4CS) software and proof-of-concept. [13] SH4CS allows the regeneration of operational software in containerised Kubernetes environments during runtime, and is able to adapt its software regeneration mechanisms based on dynamic changes in the (external) threat level. Furthermore, the UK-based Alan Turing Institute develops autonomous cyber defence technologies based on the adoption & operation of autonomous (AI-based) agents, commissioned by the UK National Cyber Security Centre (NCSC). [14] They also provide sound definitions, reasoning and research on the levels of autonomy (e.g. task autonomy or conditional autonomy) to be achieved in autonomous agents and systems.

While several organisations are internationally pushing for suitable definitions, frameworks and implementations of autonomous cyber resilient or cyber defence systems, most knowledge on this matter is still conceptual and abstract. This leaves the design and implementation of cyber resilient systems as an exercise to the interested, motivated and above all *capable* reader. Currently available implementations and technology are limited, both in number and in type of resilience functionality. Several commercial solutions, for instance, focus on checking and protecting communication on the datalink or network layers, but do not offer resilience mechanisms within operational technology (OT), endpoints or applications. [12, 15] Moreover there are, to our knowledge, no autonomous cyber resilience implementations openly available that dynamically take into account business or mission priorities, the external threat level or the current operational state of the system to-be-protected. In the following sections, we therefore not only define overarching design principles for Autonomous Cyber Resilient Systems, but also showcase an example for their hands-on implementation in TNO’s tangible use case application “OpPic”.

## Generic ACR design principles

While the need for autonomous cyber resilience mechanisms in complex digital systems is growing, generic design principles for their development and implementation are currently still lacking. This section therefore aims to provide suitable design principles and engineering inspiration to create autonomous cyber resilient products, systems and infrastructures. Our ACR principles and guidelines are intended to provide a broad basis for early-stage prototypes and proofs of concept. They are aimed to be used across domains and sectors, while leaving room for customisation to sector- and application-specific requirements.

Building on hands-on experience with prototypes as well as existing guidelines for engineering *non-autonomous* cyber resilient systems [10] and autonomic computing [11], we have identified six overarching design principles for *autonomous* cyber resilient systems:

1. Mission- and risk-awareness
2. Risk adaptive functionality
3. Nested defence
4. Periodically secure
5. Proportional response
6. As autonomous as possible

The first four principles are focused on automating cyber resilience. The last two design principles are focused on the “autonomous” aspect of ACR. Regarding authentication and authorisation, ACR principles can be quite intuitively be combined with zero-trust principles (i.e. never trust, always verify). However, in this paper those specific security aspects are left out of scope.

### 1. Mission and risk awareness

The definition of ACR – to be resilient to cyber-attacks and failures to ensure mission or business objectives without being controlled directly by humans – implies that the system needs to be mission- and risk-aware. The system should be capable to track whether it behaves as it was designed and intended to behave. Which parameters can be used for describing and measuring a system’s intended and real behaviour, and which boundaries to set is system-specific. In any case, such behavioural 'self-description' parameters should be defined in the system's design-phase, including the design of mechanisms to measure them in operation. During run-time, system behavioural parameters will then be monitored, and the monitored values will be summarised by the system's so-called 'health-state'. In addition, an adjustable set of rules should be specified that prescribes how (changes in) parameter measurements affect the system's health. These rules need to specify boundaries (e.g. threshold values) for the health-state, as well as cyber resilience actions that should be executed in case the health exceeds its boundary behaviour. Boundaries provide a way to express the system owner's 'risk-appetite' and should thus be configurable by the owner.

Note, that the mechanism of measurement, rule-based health tracking and boundary checking, which results in execution of resilience actions, is essentially a mission- and risk-aware MAPE (or OODA) loop. For some systems, it is possible to implement this MAPE loop with (relatively) generic software, such as TNO's SH4CS (self-healing for cyber security) software, that can be applied after system deployment, regardless of the exact functionality of the system to be protected. [13] For most systems however, ACR requires tweaking to, or full integration with the system's functionality which needs to be developed by-design.

## 2. Risk adaptive functionality

The potential cyber resilience actions that can be developed and executed are diverse. Using the MITRE CREF framework [18] as a basis, inspiring techniques to design automatic/autonomous risk-adaptive functionality include e.g. 'adaptive response' and 'dynamic segmentation'.

**Adaptive response**

This MITRE CREF technique includes three approaches: dynamic reconfiguration, dynamic resource allocation and adaptive management. These approaches inspire the development of non-intrusive ACR actions such as graceful degradation and risk- (or 'health'-) adaptive (de-)activation of system functionalities. In case a system's operational functionality needs to be continued under all circumstances, then any risk-adaptive actions may need to be complemented by a byzantine fault tolerant implementation of the system or the use of canary deployment (i.e. spin-up new functionality before switching off the unhealthy system).

**Dynamic segmentation**

Using dynamic segmentation and isolation as a basis, ACR systems should be able to isolate resources or to change the configurations of enclaves or protected segments, while minimising operational disruption. The design of (standardised) ACR API's or proxies for the system's interactions to other systems enables the development of risk-adaptive segmentation and isolation.

## 3. Nested defence

ACR mechanisms can be applied on different levels in systems of systems. NIST [16] defines systems of systems (SoS) as systems of interest whose elements are themselves systems. Typically, SoS with multiple heterogeneous distributed systems address large-scale interdisciplinary functionalities. SoS can be of different types based on their characteristics, such as operational and managerial independence, geographical distribution, evolving and emergent behaviours. [17] The nature of SoS (e.g. how they evolve and are under constant change) strongly increases the complexity and challenges to secure them. When systems and SoS are working together, emergent behaviour is attained: the behaviour of systems interacting with each other may be fundamentally different to that of a single independent system or component.

The abovementioned aspects make the landscape of system of systems diverse and complex, giving rise to several challenges for implementing autonomous cyber resilience. Cyber resilience needs to be considered on all levels, from each component to sub-systems, systems and system of systems. (Autonomous) cyber resilience on the component level doesn't automatically guarantee resilience at the larger system or system of systems level, and vice versa. Hence a multi-dimensional approach is required to deal with these challenges. Ensuring cyber resilience on all levels brings in the challenges of varying proportions and different scopes. In *Figure 1–3* the nesting concept for (autonomous) cyber resilience is illustrated for designing a system, a system of systems and an ecosystem of (not necessarily cooperative) systems. In addition, examples of relevant MITRE CREF techniques are referenced per level.

Another distinction between ACR functionalities at different nesting levels is that, in principle, ACR mechanisms should not change the mission objectives of the protected system itself. Recall that the ACR implementation of a system is a kind of MAPE loop tuned to a particular mission. Changing the mission would therefore invalidate the system's ACR implementation. However, at SoS level, a lower level component's or subsystem's mission may be changed in order to guard the objectives of the overall SoS (e.g. by reinstalling a new set of health- and risk parameters and new ACR rules for the adjusted mission of the subsystem).

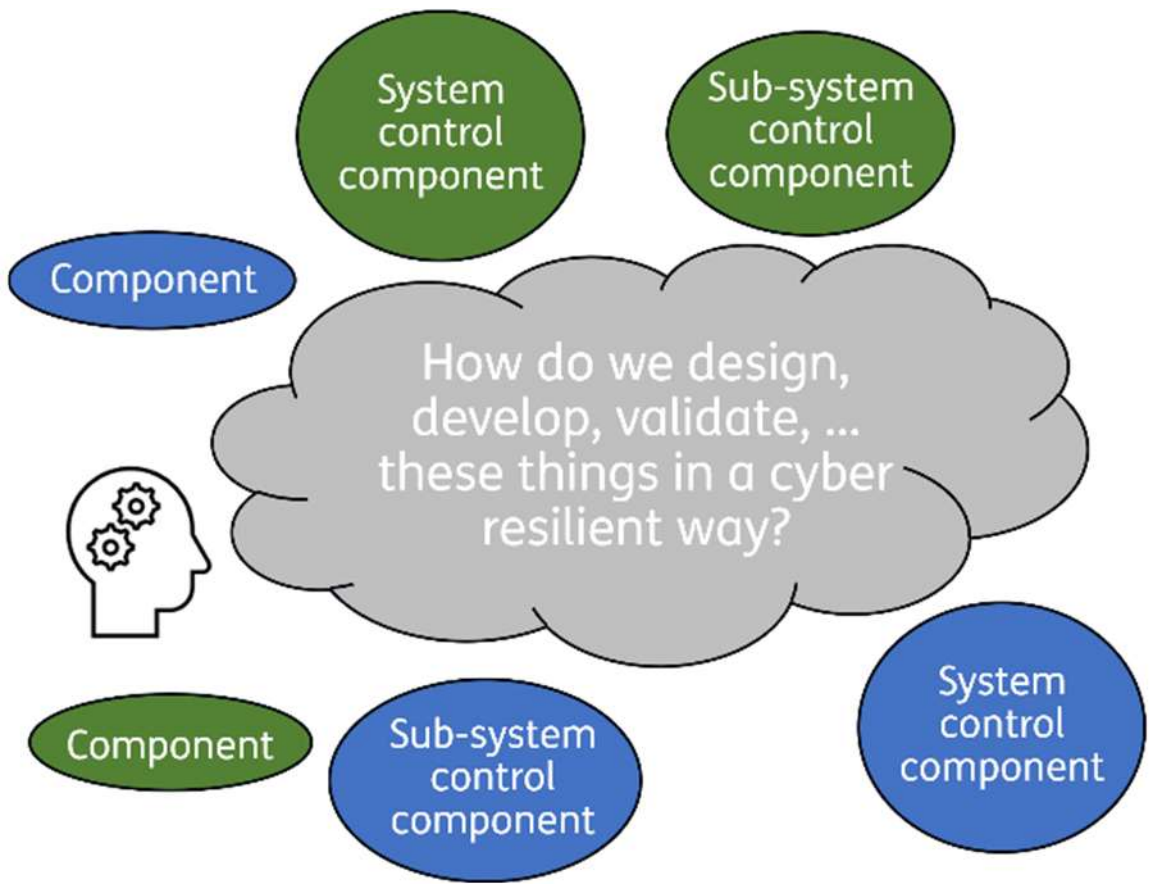

***Figure 1: <u>Creating</u> an autonomous cyber resilient product or system. Involved: 1 responsible party (the system developer).*** Example of ACR measure: Spare-capacity to ensure that a product/system can continue providing its functionality, under high(-er than usual) workload conditions. [MITRE CREF: Redundancy => Surplus capacity]

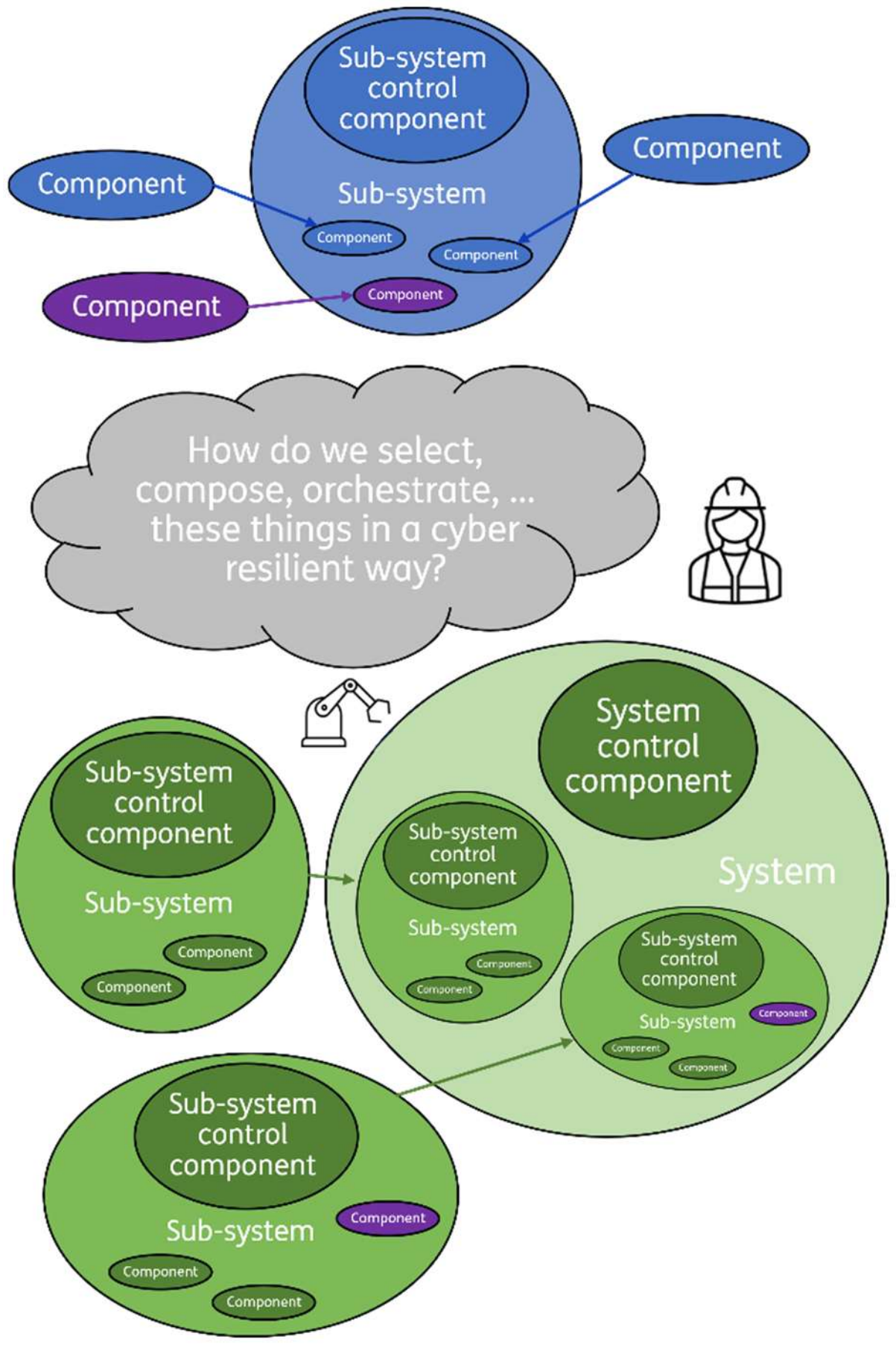

***Figure 2: <u>Composing</u> an autonomous cyber resilient system of systems from a set of autonomous cyber resilient and NON-resilient systems. Involved: 1 responsible party with several different suppliers.*** Example of ACR measure: design SoS in which multiple, functionally (almost) identical systems are available that can each provide the full SoS functionality. SoS redundancy requires a control component that governs which of the systems provides the functionality at any point in time, e.g. load-balanced active-active/passive or byzantine fault-tolerant / voting strategy. [MITRE CREF: Redundancy => Replication]

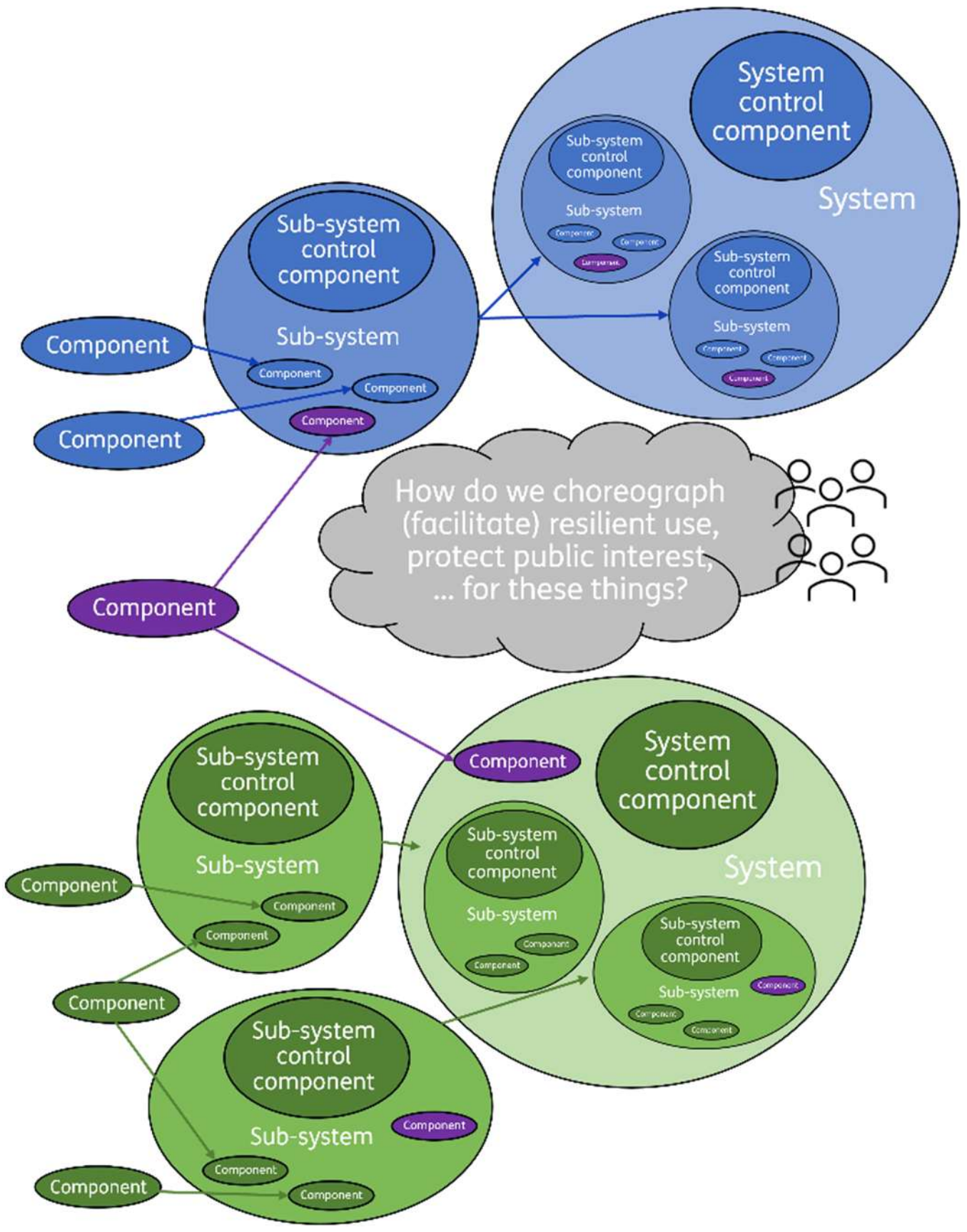

***Figure 3 <u>Assuring</u> that autonomous cyber resilient SoSs interoperate to achieve resilient behaviour. Involved: Ecosystem of several partially responsible parties with different suppliers.*** Example of ACR measure: 'Polydomy', i.e., the use of alternate SoSs with similar functionality. E.g., enabling mobile network users of provider X to use provider Y's network services, during a long outage of X's network. This requires standardisation, removing switch-over hurdles, etc. [MITRE CREF: Adaptive response => Adaptive management]

### "Thinking fast and slow" in nested SoS

In infrastructures (i.e. systems of systems) which are dynamic and emerging, making decisions on static models can lead to errors or delays with consequences to mission, business and operational objectives. Therefore, smart and adaptive decision-making is necessary. Here, we again look to biology for inspiration: According to the psychological concept "Thinking fast and slow" [18], human decision making in the brain is essentially performed in two control loops: The first one is a fast loop that automatically executes quick responses to a trigger intuitively or rule-based, without much analysis. The second loop, in contrast, can construct thoughts in an orderly and logical series of steps, but is much slower as a result.

Applying the "thinking fast and slow" concept in digital IT systems from a cybersecurity perspective, a central Security Operations Centre (SOC) can be compared to the "slow loop" of human decision making: the SOC is capable of making decisions based on aggregated data following sequential steps and processes. Considering the vast amount of data to be processed and analysed, this is by default a slow and resource-intensive process. During a cyber-security incident, a SOC team would need to make hundreds of reactive decisions. In comparison, applying ACR techniques (e.g. self-healing and self-protection mechanisms) locally to components and sub-components of the infrastructure would be equivalent to a "fast loop" reaction: quick, intuitive (rule-based) and automatic. This process can be improved using autonomous (AI-) agents that adapt to previously unknown conditions as described by Lohn et. al [19], who proposed characteristics for autonomous cyber defence agents.

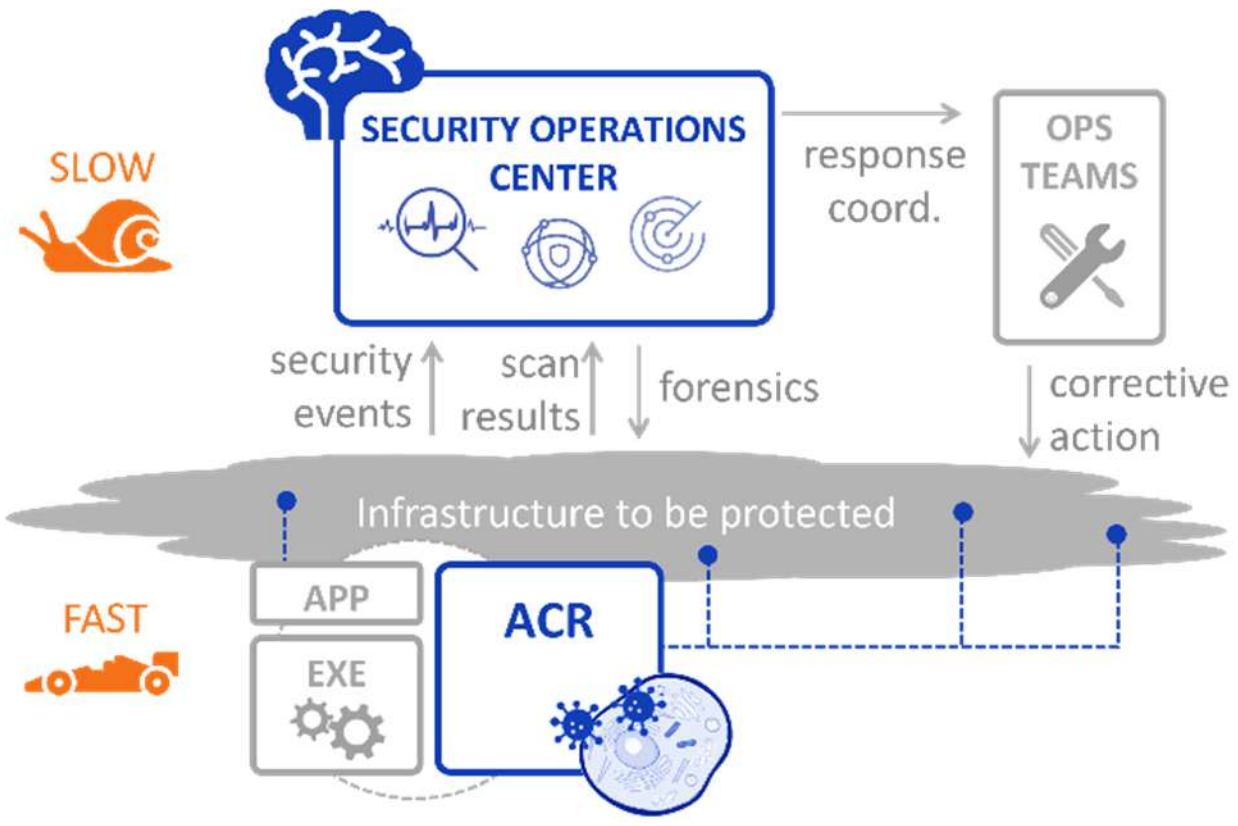

*Figure 4: Schematic of the "fast" and "slow loop" of human decision making, applied to cyber resilience functionality.*

## 4. Periodically secure

A powerful measure for restoring software components of a system from problems such as software failure, cache poisoning, in-memory malware etc. is to kill and restart the system. This is also one of the reasons why the CRA[1] requires that digital systems can be reset to their original state. However, the ACR principle is much stronger than the CRA requirement: *ACR systems shall be able to periodically reconfigure themselves into a state that is 'accredited secure'.* The vendor of an ACR system needs to specify measurable system conditions for which it declares that the system is secure. One such "accredited secure" condition is the system's condition when taken out-of-the-box (i.e. the original state).

A possible method for periodically bringing an ACR system into an 'accredited secure' state is to use the ACR MAPE loop of measurement, rule-based health tracking and execution of resilience actions. Specifically, 'accredited secure' states need to be defined as states in which the system health is optimal. Creating a system that is periodically brought into an accredited secure state requires to specify a rule that periodically checks whether the optimal health state has been reached in the past period. If this is not the case, it will restart the software component of the system. Such a 'regeneration' mechanism can in principle be implemented for any product or system that complies to the CRA. Vice versa, a periodically regenerating system automatically complies with the related CRA requirement.

Implementing such a regeneration mechanism brings more benefits than just CRA compliance and early termination of (known or unknown) malicious behaviour. For example, it suggests a natural implementation mechanism for automated patching: the system's software may be restarted/ regenerated from the original software version, but by appending the mechanism with a DevSecOps pipeline the regeneration action can just as easily be done using a patched software version (which is accredited secure by the vendor). A DevSecOps deployment pipeline is an example of an ecosystem-level ACR measure. Another benefit of regeneration is that it provides an additional way to incorporate risk- adaptiveness in ACR systems: under some circumstances, indicators of compromise may lead to the conclusion of a raised risk level that does not require immediate action. In such circumstances, it can be desired to increase the frequency of health status verification and/or reduce the time until the next periodic regeneration (or all subsequent ones). This results in a situation where the ACR system gets into a higher alert phase.

These examples illustrate how the principle of "periodically secure" is a fundamental ACR building block for adding additional cyber resilience features.

## 5. Proportional response

ACR mechanisms should not only be mission- and risk-aware, but their responses also need to be

[1] Cyber Resilience Act, see [8].

*proportional* to any business- or mission risk posed by cyber-attack or resulting operational failure. Cyber resilience principles are intended to reduce and contain risk while safeguarding business continuity, which depends on the available operational resources. Both "overreactions" and a lack of suitable action can harm the operations of the digital system to be protected. Examples of implementing proportional response is reducing the duration of regeneration periods under high-risk circumstances (e.g. when an attack/infection is likely). Vice versa, the time period between regenerations may be increased (up to some maximum) under low-risk circumstances so as to not claim unnecessary system resources and maximise system availability and uptime.

Another part of "proportional response" is the concept of *cyber response suppression* in analogy with suppression mechanisms in the immune system, which aims to increase survival chances in exceptionally stressful situations. For instance, the immune system of an antelope running from a lion gets temporarily suppressed to increase the antelope's overall survival chances – more attention and energy are allocated to the running task, as opposed to the task of fighting a potential viral infection. In digital systems, such a high-risk response suppression (e.g. keep an important radar system running despite it likely being compromised) may get automated with thoroughly developed rule-sets, but may also be implemented through human override functionality.

To avoid ACR under- or overreactions that may negatively impact system functionality, it is imperative to perform risk-analyses and rigorous testing of ACR methods during the design phase.

### 6. As autonomous as possible

An ACR system's *level of autonomy* must be proportional to the owner's risk-appetite and (possibly) the current health level. To maintain ongoing operations as much as possible, organisations may start with a low level of system autonomy when carrying out specific tasks, and increase the level of system autonomy over time and with the system's growing capabilities for situational awareness, keeping check on operational states and safeguarding business/mission priorities.

Challenges faced with autonomous systems, have already given rise to several frameworks and publications such as those by NIST, the Alan Turing Institute and others [14, 20, 21]. In these publications, different levels of system autonomy are defined, while at the same time considering the system's operational objectives. Already in the design phase, system developers will need to define clear thresholds and mechanisms to determine levels of autonomy based on risk-appetite and health levels. The (semi-) autonomous functionality then needs to be developed and implemented based on the available functions of the system. To illustrate the guideline of proportional autonomy, we give an example of three different autonomy levels and their configuration requirements below. Note that, if desired, additional (intermediate) levels may always be added to suit a certain organisation's business or mission need.

**Autonomy level: Low**

- *Capability:* Limited ACR capabilities. System is predominantly managed by the human operator.
- *Configuration requirements:* Mainly logging functionality, without ACR initiated interaction to the human user or security specialist. No autonomous actions are allowed to be taken that could interfere with operations.

**Autonomy level: Medium**

- *Capability:* Partially autonomous systems continue to operate on essential mission and business objectives with limited human intervention.
- *Config requirements:* Some autonomous actions are allowed, such as limited configuration changes. Several actions require the human-in-the-loop to be informed and to approve of the action.

**Autonomy level: High**

- *Capability:* System has sufficient situational and business/mission awareness to be capable of high-stake decisions while completing functional objectives without human intervention. This system is also able to autonomously adapt to dynamic conditions while maintaining operation.
- *Config requirements:* Autonomous actions are allowed and implemented. Autonomous reasoning capabilities can perform suitable changes to the system configuration, or to any pre-defined rule sets for fast regeneration or response.

## Illustration of ACR design principles

Previous self-healing proofs of concept provide basic demonstrations of ACR principles in cloud environments. [13] In such environments, features such as containerisation, monitoring and easy deployment are natively supported. As a result, the development of stateless, 'disposable'[2] applications that are autonomously cyber resilient is relatively straightforward: due to the stateless nature and availability of redundant resources, even a 'harsh' resilience mechanism such as kill-and-restart has only limited negative impact on system resources and availability. As a result, the response is easily proportional to the threat, the level of autonomy can be relatively high, and regeneration may even be accelerated/intensified with decreasing health level. Moreover, in the cloud, autonomous cyber resilience can be implemented on the deployment level without having to integrate self-healing functionality with application functionality. [13]

However, developing ACR for more complex digital systems is still an intricate challenge. For systems that are stateful, or for which the risk level (of some) of its functionality may vary strongly in time, implementations of ACR will have to be integrated with system functionality. Digital components in critical naval and aeronautical systems are typical examples of such systems. To illustrate how the ACR design principles can be applied to a more complex digital system than a Kubernetes cloud environment TNO developed the "OpPic" application, inspired by naval radar systems and Air Traffic Control.

## OpPic

OpPic is a vehicle manoeuvring application, that implements a primary / alternate mode for functional resiliency purpose[3]. OpPic consists of four components that are implemented as separate containers in a Kubernetes platform (see Figure 5):

a) A client that provides the user with an 'operational picture' and actuators to manoeuvre the vehicle left, right, faster, slower;
b) A primary (PRI) and alternate (ALT) OpPic component that process route and manoeuvring data into an operational picture;
c) A route data generator that generates coordinates of the target route up to the horizon, in a pace determined by the speed that is controlled by the user.

The client consumes operational pictures produced by the PRI and ALT components, and only displays output from the component that is selected by the user. In real-world applications it is common that the functionality of the PRI component is very rich (e.g. using and contextualising inputs from many different sources), while the ALT component features more concise functionality (e.g. using only one filtered input source). The primary component is then typically the user's preferred selection, but it is also more vulnerable (e.g. to malicious data inputs or data-driven errors). As a result, the user will typically select the more feature-rich PRI during run-time, unless under some specific (adversarial) circumstances only the ALT is able to provide a stable operational picture.

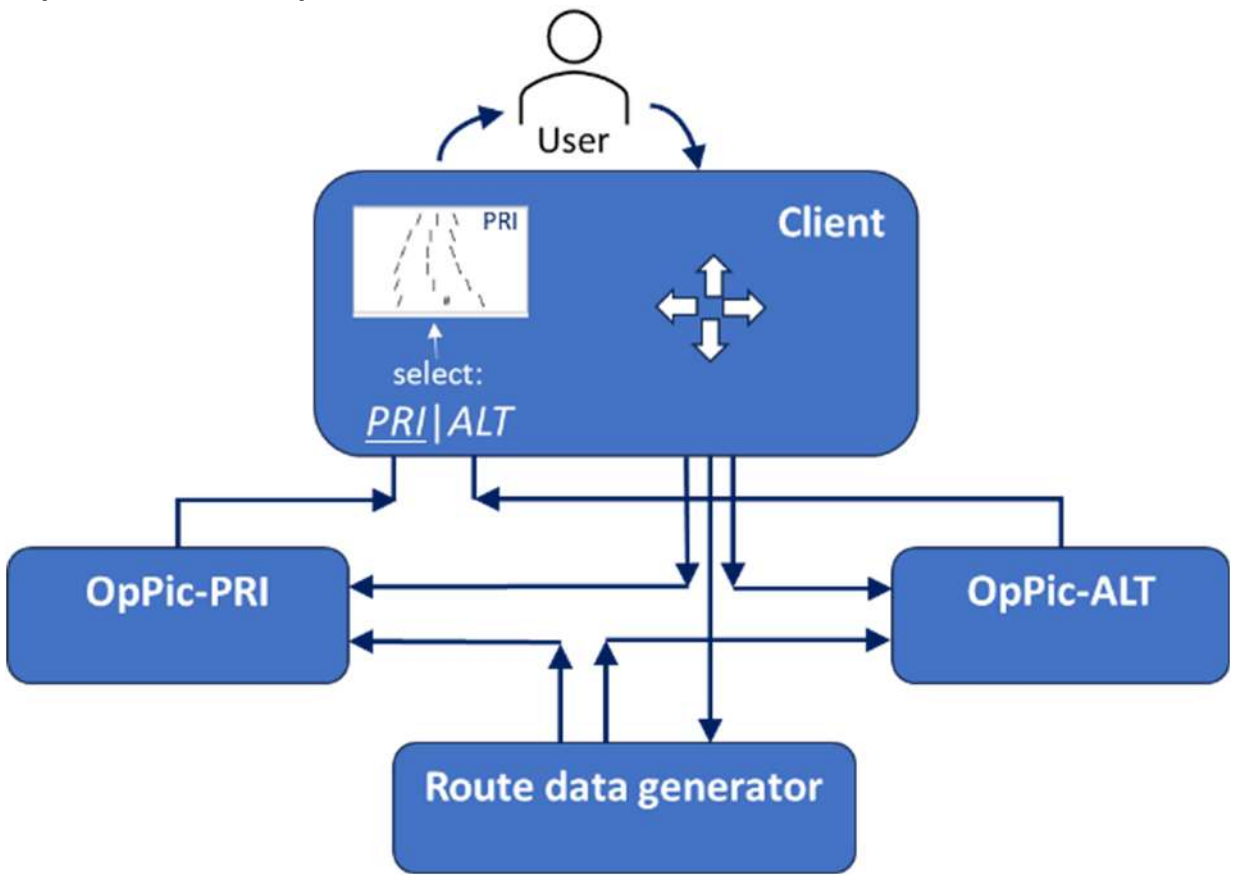

*Figure 5 OpPic components and interactions.*

## OpPic resilience

The PRI/ALT modes provide common user-coordinated (non-autonomous), application level resilience. This resilience can be improved by adding ACR principles. In particular, we add the self-healing and self-protection feature of software container regeneration that was introduced in [13] (i.e. to implement design principle 4: periodically secure). In short, container regeneration ensures that in a Kubernetes environment any container is periodically stopped and restarted from a clean container image. This mechanism prevents (long-time) persistence of undesired cache poisoning, memory leaks and/or overflow that can lead to malicious code execution, data-driven and runtime errors and performance degradation.

However, while container regeneration for stateless, disposable web server replications could in earlier

[2] Applications can be 'disposable' for multiple reasons. For example, because the application may be implemented using multiple identical instances (e.g. web server front-ends) or the functionality provided by the application is not essential.

[3] A generally applicable resiliency method for critical applications is PACE [22]. It describes how application functionality can stepwise be scaled down during incidental period of unavailability. For the OpPic application, it suffices to illustrate PACE's Primary and Alternate modes; including the Contingency and Emergency modes is conceptually 'more of the same'.

ACR PoCs be executed 'at will', OpPic's stateful nature makes it obvious that a more careful approach needs to be taken. In particular, the second ACR design principle prescribes that ACR should be risk-adaptive and the fifth principle prescribes that ACR actions should be risk-proportional. As explained earlier, the level of ACR autonomy must be tuned to the application (sixth principle). For OpPic it makes sense to perform container regeneration of PRI only after a graceful switch-over to ALT, e.g. by presenting a count-down pop-up to the user (which can be overruled with 'execute now' and/or 'postpone' buttons), in order to keep the user in full charge of PRI / ALT selection.

Further, presuming that the ALT component is selected by the user in case the PRI is unavailable or unreliable, a situation where ALT is used may be regarded as an indication of high risk. In such a case the ALT container regeneration should be postponed.

### Executing ACR logic

In TNO's earlier SH4CS prototype for container regeneration for stateless, 'disposable' containers, the MAPE control loop was a simple, static flow chart [13]. Implementing ACR for more complex systems requires a different approach. To implement ACR for OpPic, an extensible list of rules is used. Each rule consists of an event (e.g. an indicator that the regeneration timer of the OpPic-ALT container has expired), a condition (e.g. the client container has selected OpPic-PRI as source) and an autonomous resilience action (e.g. OpPic-ALT container is regenerated). Similar to the earlier SH4CS prototype, an ACR container sidecar[4] is running along each OpPic application container. This sidecar periodically checks for triggered events and executes actions according to predefined rule-sets, if the logical condition is satisfied. This simple rule-based logic suffices for implementing the described risk-conditional container regeneration example. The extensibility of the rule-base makes ACR sidecars adaptable, for example to create nested ACR loops where the sidecar rules are updated by a (possibly AI-based) 'thinking slow' MAPE loop.

### Lessons learned from OpPic

The design of OpPic with ACR features involves many considerations, such as constraints on when the PRI container may regenerate, and how periodic regenerations cannot be indefinitely postponed (design principles: risk proportionality and level of autonomy). It is beyond the scope of this paper to present an exhaustive list of all design considerations. The presented examples are merely illustrating how our generic ACR design principles can be used in the OpPic example and in the design of more complex systems. Moreover, OpPic makes plausible why – for more complex digital systems – ACR needs to be considered during the design phase as a functional requirement. After all, container regeneration requires availability of an interface to OpPic's client state information such as the selected mode. ACR is not a non-functional aspect, as opposed to the common (historical) understanding that traditional cybersecurity is non-functional.

In this paper the third design principle of nested control loops is not explicitly addressed, nor is the concept of exchanging threat-level information between ACR sidecars. This paper also does not go into secure techniques for sharing ACR information between sidecars (where the principle of zero trust plays an important role). In this exploratory phase of ACR, such promising concepts are yet to be investigated in more detail.

## Conclusions

In this paper, we have addressed both the necessity and practical challenges of introducing autonomous cyber resilient complex digital systems. To aid the design of future-proof ACR systems, we have shared a first definition of six ACR design principles and illustrated their use (alongside with remaining open questions and challenges) for the example of TNO's OpPic application. With OpPic, we demonstrate how ACR self-healing and self-protection mechanisms may be practically integrated with (stateful) application level resilience mechanisms that are typically present in military or aeronautics systems.

While this exploration is a first step in moving towards autonomous cyber resilient system design and implementation, many challenges are still to be solved for a large transition and important paradigm shift to take place: autonomous cyber resilience by design. Work is currently ongoing on more thorough definitions of ACR design principles, as well as implementations of ACR proof of concepts in complex real-life applications.

[4] In Kubernetes the term sidecar is used for containers that 'run along' with an application container to extend its functionality, e.g. for cybersecurity purposes.